\documentclass[aps,prl,reprint,groupedaddress]{revtex4-2}

\usepackage{physics}
\usepackage{graphicx}
\usepackage{xcolor}

\begin{document}

\title{High-rate sub-GHz linewidth bichromatic entanglement source for quantum networking}

\author{Alexander N. Craddock}
\author{Yang Wang}
\author{Felipe Giraldo}
\author{Rourke Sekelsky}
\author{Mael Flament}
\author{Mehdi Namazi}
\email{Corresponding author. mehdi@quconn.com}
\affiliation{Qunnect Inc., 141 Flushing Ave, Ste 1110, Brooklyn, NY 11205-1005}

\date{\today}

\begin{abstract}
The generation of entangled photon pairs which are compatible with quantum devices and standard telecommunication channels are critical for the development of long range fiber quantum networks. 
Aside from wavelength, bandwidth matching and high fidelity of produced pairs are necessary for high interfacing efficiency.
High-rate, robust entanglement sources that satisfy all these conditions remain an outstanding experimental challenge.
In this work, we study an entanglement source based on four-wave mixing in a diamond configuration in a warm rubidium vapor.
We theoretically and experimentally investigate a new operating regime and demonstrate an entanglement source which produces highly non-degenerate $795$ and $1324$-nm photon pairs.
With this source we are able to achieve in-fiber entangled pair generation rates greater than $10^7\, /s$, orders of magnitude higher than previously reported atomic sources. 
Additionally, given our source's native compatibility with telecom infrastructure and atomic systems, it is an important step towards scalable quantum networks.
\end{abstract}

\maketitle

\section{Introduction}

Entanglement is a cornerstone of the field of quantum information.
Within the field, a number of applications depend on the ability for two or more parties to share entanglement over long distances, including distributed quantum computing \cite{monroe_large-scale_2014}, distributed quantum sensing \cite{eldredge_optimal_2018}, and quantum communication \cite{ekert_quantum_1991}.
Entangled photons are the natural candidate for realizing long-distance entanglement distribution.
However, owing to optical losses in quantum channels the implementation of quantum repeating \cite{Duan2001} will be necessary.
An integral part of many quantum repeating protocols is the efficient interfacing of entangled photons with quantum memories \cite{sangouard_quantum_2009}, in addition to various quantum computing and sensing platforms.
Robust sources of entangled pairs of photons which are simultaneously compatible with quantum devices and standard quantum channels (e.g., telecom fibers) are therefore likely to be an necessary component in the realizations of a quantum internet. 
Aside from producing appropriate wavelengths, a practical  entanglement source for general purpose quantum networking must satisfy multiple criteria simultaneously, such as high rate generation, narrow line-width photons, high fidelity, and heralding efficiency.
However, the construction of such sources remains an outstanding experimental challenge.

Given their compactness, robustness, simplicity, and native compatibility with atom-based quantum memories, warm-atomic ensembles are a natural system in which to implement entanglement sources.
Much of the recent literature on the topic has focused on four wave mixing (FWM) in ladder \cite{davidson_bright_2021,lee_highly_2016,davidson_bright_2023,park_polarization-entangled_2019,park_direct_2021,ding_generation_2012} or double $\Lambda$ configurations \cite{chen_room-temperature_2022,hsu_generation_2021,shu_subnatural-linewidth_2016,zhu_bright_2017,mika_high_2020,jeong_temporal-_2020}.
The ladder scheme has been used to realize high absolute brightness (pairs per unit time) sources, $\approx 4\times10^5\, /s$ \cite{davidson_bright_2021}, while the double $\Lambda$ configuration has been used to demonstrate high spectral brightnesses (pairs per unit time per unit bandwidth), $\approx 4\times10^5/s/\mathrm{MHz}$ \cite{chen_room-temperature_2022}.
However, the high performance of these sources is typically restrictive on the wavelengths of the photons produced, which are in the near-infrared (NIR) and are thus unsuitable for long-distance entanglement distribution across telecom fiber networks.
Additionally, demonstrations of entanglement in these systems are limited \cite{park_direct_2021, park_polarization-entangled_2019} owing to the near-degeneracy of the transitions.
Work also exists on warm-atom FWM in diamond configurations \cite{willis_correlated_2010,willis_photon_2011}.
These sources have been shown to natively produce entangled photon pairs with disparate wavelengths, allowing them to be simultaneously compatible with atomic quantum memories \cite{wang_field-deployable_2022} and telecom fiber networks.
However, they have typically been less performant than their ladder and double $\Lambda$ counterparts, with demonstrated peak absolute and spectral brightnesses of $\approx 5\times10^3/s$ and $\approx 5 /s/\mathrm{MHz}$ respectively \cite{willis_photon_2011}.
This has been attributed to the inability to simultaneously address all the velocity groups within the vapor in the photon production process \cite{lee_highly_2016}.

\begin{figure*}[t]
    \centering
    \includegraphics[width=1\linewidth]{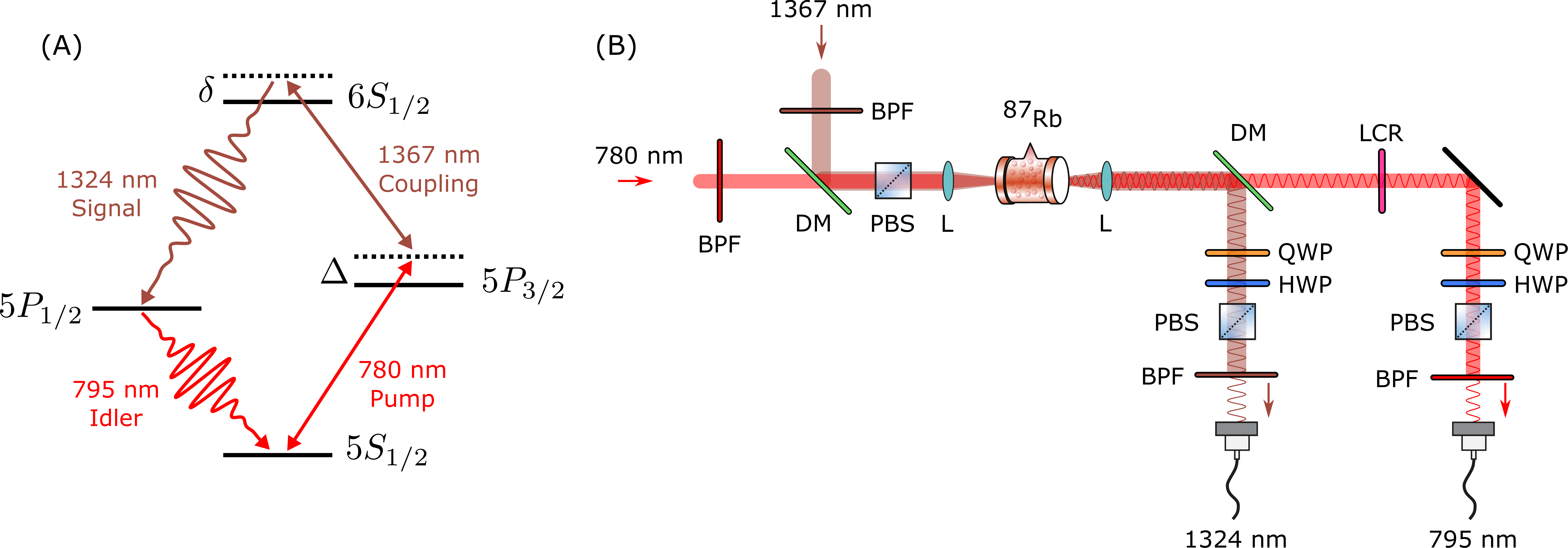}
    \caption{(A) Relevant rubidium level diagram for the four-wave mixing process.
     (B) Experimental setup for generation and analysis of entangled photon pairs.
    PBS: polarizing beam splitter, DM: dichroic mirror, BPF: bandpass filter, LCR: liquid crystal retarder, QWP: quarter waveplate, HWP: half waveplate, L: Lens.}
    \label{fig:experimental_setup}
\end{figure*}

In this paper, we demonstrate a source of bichromatic entangled photon pairs based on diamond-scheme FWM in a warm rubidium ensemble with both high absolute and spectral brightness.
We theoretically and experimentally explore a new operating regime in the diamond scheme that allows us to address all the atoms within the vapor and achieve pair rates in-fiber exceeding $\approx 10^7/s$, and a lower bound on the $\ket{\Phi_+}$ Bell state fidelity of $>95$\%.
To the best of our knowledge this is the highest demonstrated absolute brightness for a warm-atom entangled photon pair source. 
Additionally, it produces telecom-NIR entangled photon pairs ($1324$ and $795$ nm respectively), making our source suitable for use in fiber-based quantum repeating applications.

\section{Theory}

\begin{figure}[t]
    \centering
    \includegraphics[width=\linewidth]{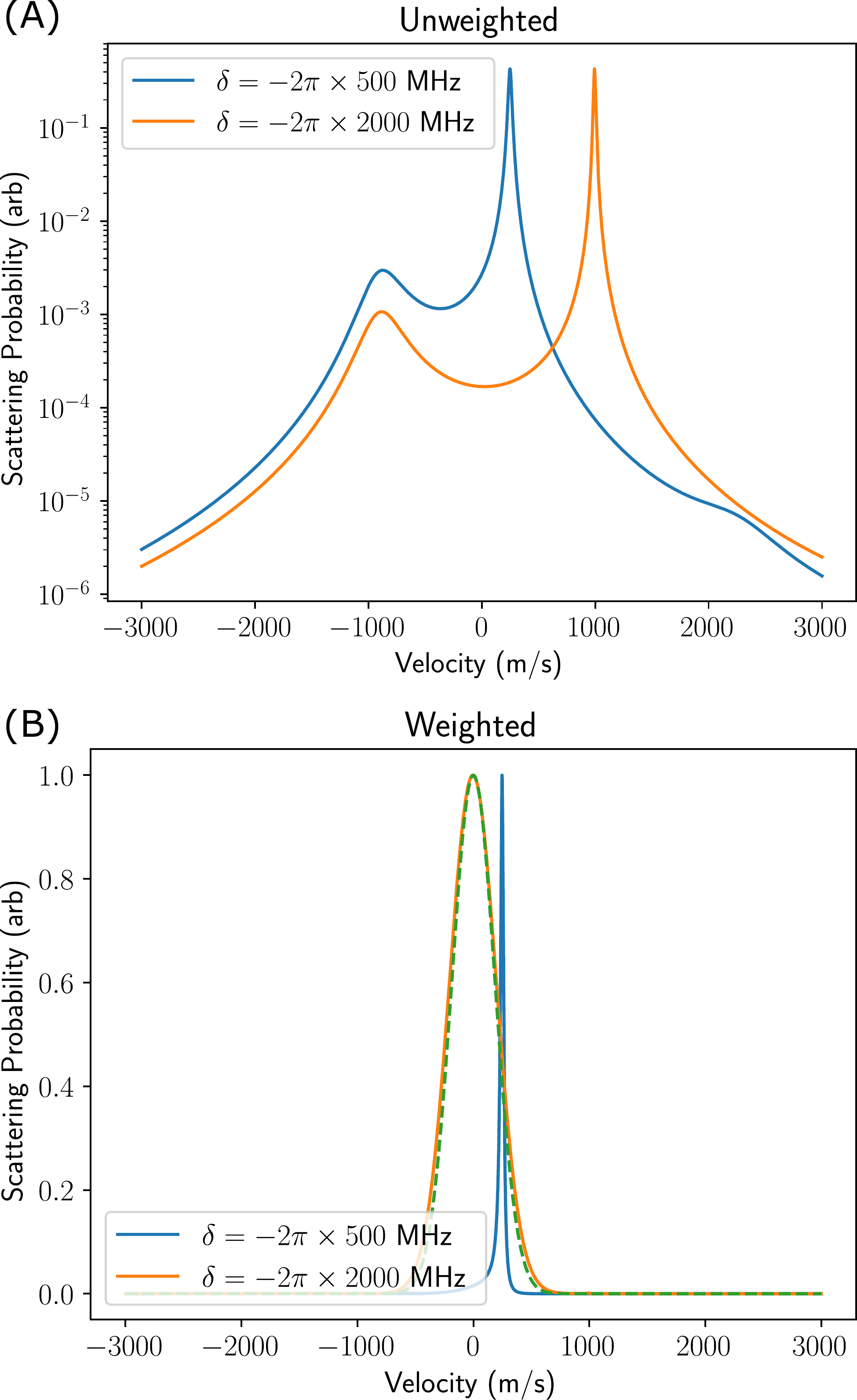}
    \caption{Numerical simulations of the simplified three-level system in the near (blue) and far-off (orange) two-photon resonance regimes.
    Plot (A) ((B)) shows the scattering probability, unweighted (weighted) by the Maxwell-Boltzmann distribution, as a function of atomic velocity.
    (B) has been re-normalized so that the peak weighted scattering probability is unity for ease of comparison.
    Dashed green line in (B) shows the Maxwell Boltzmann distribution for vapor temperature used in the simulation.
     In both cases the pump detuning is fixed far from resonance ($\Delta = 2\pi\times 1150$ MHz), the pump and coupling Rabi frequencies are held at reasonable experimental values ($\Omega_{p}=\Omega_{c}=2\pi\times 350$ MHz), and temperature is fixed at $T=80^\circ$ C. }
    \label{fig:Theory}
\end{figure}

In this work we explore spontaneous FWM in a warm ${}^{87}$Rb vapor.
The relevant atomic level structure for the diamond FWM process is shown in Fig. \ref{fig:experimental_setup}(A).
To understand the physics of the excitation process we analyze the three level Hamiltonian, under the rotating wave approximation, for an atom with velocity, $v$
\begin{equation}
    \hat{H}(v)=\hbar
    \begin{pmatrix}
        0 & \frac{\Omega_{p}}{2} & 0 \\
        \frac{\Omega_{p}^*}{2} & -\Delta(v) & \frac{\Omega_{c}}{2} \\
        0 & \frac{\Omega_{c}^*}{2} & -\delta(v)
    \end{pmatrix},
\end{equation}
in the $\left\{\ket{5S_{1/2}},\ket{5P_{3/2}},\ket{6S_{1/2}}\right\}$ basis.
We treat the $\ket{6S_{1/2}}\rightarrow\ket{5P_{1/2}}\rightarrow\ket{5S_{1/2}}$ as an effective $\ket{6S_{1/2}}\rightarrow\ket{5S_{1/2}}$ decay, in addition to the $\ket{6S_{1/2}}\rightarrow\ket{5P_{3/2}}$ and $\ket{5P_{3/2}}\rightarrow\ket{5S_{1/2}}$ decay channels.

For ease of calculation we treat atomic motion as one dimensional along the beam propagation direction.
This is a valid approximation for our experimental setup (see experimental setup section).
Additionally, to avoid significantly populating the $\ket{5P_{3/2}}$ state we use a large ($\Delta = 2\pi\times 1150$ MHz) single photon detuning.
We numerically solve to find the steady state for the Liouvillian associated with the reduced three-level system.
As the $\ket{6S_{1/2}}\rightarrow\ket{5P_{1/2}}$ decay rate is proportional to the $\ket{6S_{1/2}}$ steady state population, we use this as a proxy for the signal photon scattering probability.
In Fig. \ref{fig:Theory}(A) we show two examples of the scattering probability as a function of atomic velocity to demonstrate the near (blue) and far-off (orange) two-photon resonant regimes.
In both regimes a sharp resonant peak ($v\approx250\mathrm{m/s}$ and $v\approx1000\mathrm{m/s}$ for the near and far detuned case in Fig. \ref{fig:Theory}(A)) is seen in the scattering probability at
\begin{equation}
    v\approx-\frac{c\delta}{\omega_{6S_{1/2}}},
\end{equation}
due to the two-photon Doppler shift, where $\hbar \omega_{6S_{1/2}}$ is the energy of the $\ket{6S_{1/2}}$ state.
Two broader and less prominent peaks are seen in the figure at $v\approx-1000\mathrm{m/s}$, where the $\ket{5P_{3/2}}$ is resonantly excited.
To determine the scattering probability in the atomic vapor we weight the $\ket{6S_{1/2}}$ population by the atomic velocity population, as given by the Maxwell-Boltzmann distribution, shown in Fig. \ref{fig:Theory}(B).
For the near detuned case a significant fraction of the population resides at the resonant velocity class, giving rise to a sharp feature in the figure.
In contrast, for the far-off two-photon resonance regime the resonant velocity has a near negligible population fraction.
When we weight by the likelihood of the atomic velocities we find that the majority of the scattering occurs off-resonantly despite the unweighted scattering probability for this process being significantly smaller than for resonant scattering.
As seen in Fig. \ref{fig:Theory}(B), for far-off two-photon resonance excitation the weighted scattering probability follows the Maxwell-Boltzmann distribution.

The significance of this behavior can be understood by considering the collective excitation projected onto the atomic system upon detection of a signal photon \cite{davidson_bright_2021}.
The likelihood of phase matched emission of the idler scales with the atom number participating in the collective excitation \cite{saffman_creating_2002}.
Given that the collective excitation has a distribution similar to that of the weighted scattering probability, we expect, for a fixed atomic density, phase matched emission of the idler to be more likely for off-resonant excitation relative to near-resonant excitation.
For a given vapor temperature, we therefore expect higher heralding efficiencies when the source is operated in the far-off-resonant regime.

We note that several groups \cite{davidson_bright_2021,lee_highly_2016} have utilized nearly-Doppler-free excitation schemes in warm vapors in order to address all velocity groups simultaneously.
However, these configurations have tight restrictions on the similarity of the wavelength of the coupling and pump beams.
The off-resonant excitation scheme presented here achieves the same effect even in systems with disparate coupling and pump wavelengths.

\section{Experimental Setup}

In our experiment we use a $780$-nm pump and $1367$-nm coupling laser to drive the two-photon $\ket{5S_{1/2}}\rightarrow\ket{6S_{1/2}}$ transition, via the intermediate $\ket{5P_{3/2}}$ state.
For this work the pump light is frequency stabilized to the ${}^{85}$Rb $\ket{5S_{1/2}, F=3}\rightarrow\ket{5P_{3/2},F'=4}$ transition, and thus $\Delta\approx 2\pi\times1.1$ GHz (blue detuned from the ${}^{87}$Rb $\ket{5S_{1/2}, F=2}\rightarrow\ket{5P_{3/2},F'=3}$ transition).
To stabilize the coupling light we use a dual resonance optical pumping (DROP) setup \cite{perez_galvan_measurement_2008}.
We use an electro-optic modulator (EOM) to allow us to offset the detuning of the $1367$-nm light from resonance.

The experimental setup is shown in Fig. \ref{fig:experimental_setup}(B).
The pump and coupling beams are overlapped on a shortpass dichroic mirror so that they co-propagate through a $5$-mm long enriched ${}^{87}$Rb vapor cell.
Both the pump and coupling beams are horizontally polarized prior to the vapor cell using a common polarizing beamsplitter (PBS).
We use a pair of $f\approx 50$ mm achromatic lenses to focus the beams to a $1/e^2$ beam radii of $\approx 30$ $\mu$m inside the cell.
We collect the $1324$-nm signal, and $795$-nm idler at zero angle to the pump, which naturally satisfies the phase matching condition \cite{willis_correlated_2010}.
Both signal and idler are coupled into collimators with equivalent $1/e^2$ mode radii of $\approx 20$ $\mu$m in the vapor cell.
We heat the cell using a pair of metal ceramic heaters placed directly on the faces of the cell to prevent condensation.

After the cell we use a shortpass dichroic mirror to separate the telecom and near-infrared light.
We use bandpass filters to suppress leakage of the coupling beam into the signal fiber, and pump leakage into the idler mode.
On both the signal and idler path, a quarter waveplate (QWP), half waveplate (HWP), and PBS allow us to select the detected polarization modes.
We align a liquid crystal retarder (LCR) on the idler path so that the slow axis is in the vertical direction.
By tuning the retardance we can apply arbitrary phase shifts between the $\ket{H}$ and $\ket{V}$ polarization modes, allowing us to compensate for differential phase shifts in the setup.

We detect the idler photons using a single photon avalanche photodiode, and the signal photons using a superconducting nanowire single photon detector ($\approx 350$-ps and $\approx 90$-ps timing jitter respectively).
The detection efficiency, which includes only the quantum efficiencies of the detectors and losses after the photons are coupled into the fibers, is $78 (2)$\% and $68$\% for the signal and idler photons respectively.
We use a time tagging device ($1$-ps temporal resolution) to log events from the two detectors, from which we process coincidences.

\section{Results}

\begin{figure}[t]
    \centering
    \includegraphics[width=\linewidth]{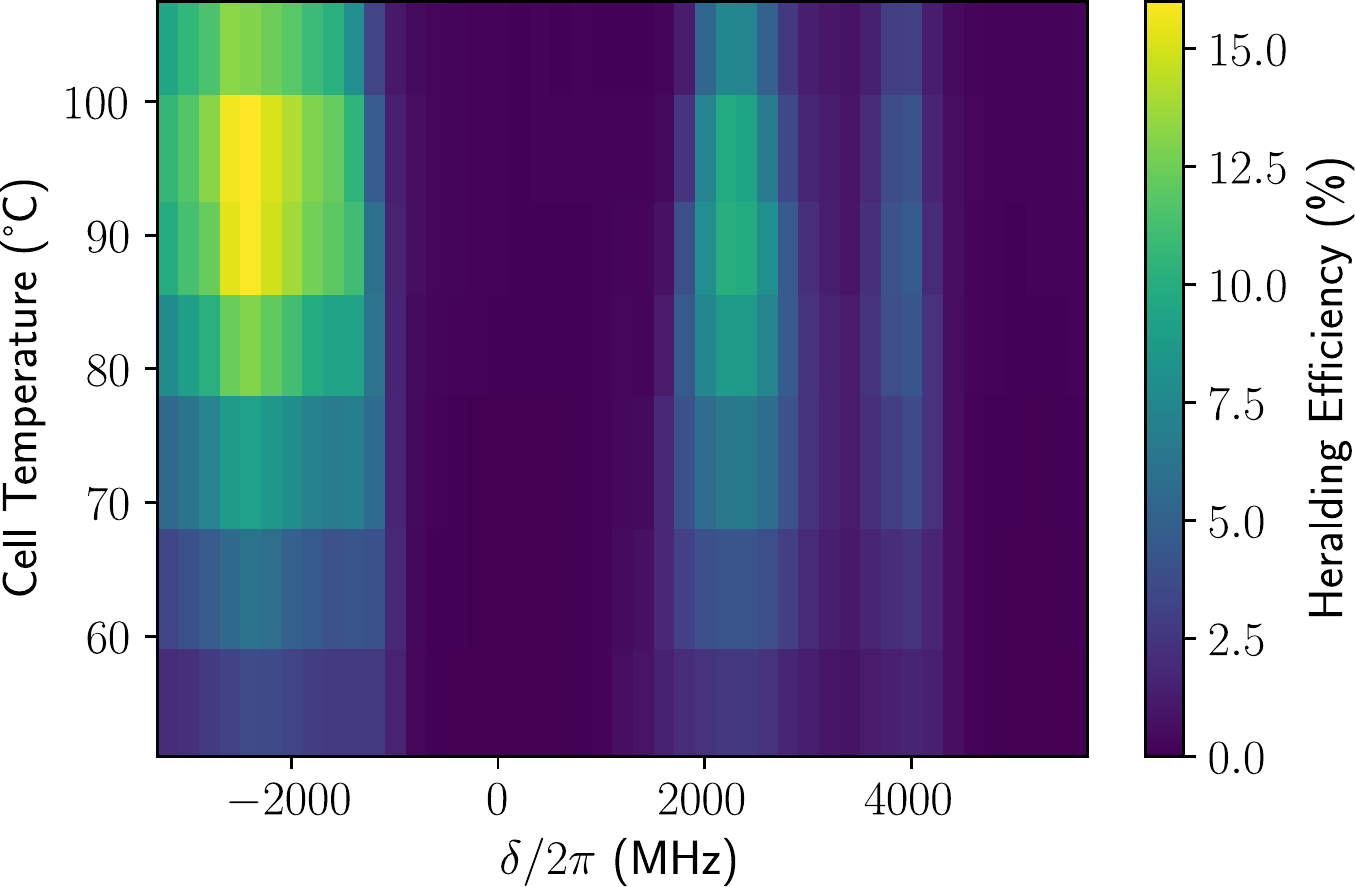}
    \caption{Heralding efficiency of the source
    as a function of the vapor cell temperature, and the two-photon detuning, $\delta$.
    The single photon detuning is fixed, $\Delta\approx 2\pi\times1.1$ GHz.
    The pump power is fixed at $\approx 250$ $\mu$W.
    For each detuning/temperature we vary the coupling power to set the signal rate to $\approx 100$ kcps.}
    \label{fig:heralding}
\end{figure}

We first explore the dependence of the source parameters on the two-photon detuning and the vapor cell optical depth (OD, measured for 795-nm idler photons).
To adjust the OD we change the vapor cell temperature, while the two-photon detuning is varied using the EOM used in the offset DROP lock.
For this data we fix the pump power at $\approx 250$ $\mu$W.
We vary the coupling power for each detuning/temperature to bring the detected signal rate to $\approx 100$ kcps.
Here, we only measure the $\ket{VV}$ mode of the source.

We look at the heralding efficiency, which is the probability of detecting a $795$-nm photon upon the detection of a $1324$-nm photon.
As we can see in Fig. \ref{fig:heralding}, for fixed atomic temperature, there is a trend that increasing two-photon detuning increases the heralding efficiency.
As explored in the theory section, this behavior is expected as large two-photon detunings address all the velocity classes simultaneously, rather than a narrow, resonant band.

From the simple theory described above we expect the source behavior to be symmetric about zero two-photon detuning.
However, we see that is clearly not the case in Fig. \ref{fig:heralding}.
We attribute this reduced heralding efficiency for positive two-photon detunings to undesired interaction with ${}^{85}$Rb atoms within the vapor ($\approx 99$\% ${}^{87}$Rb vapor cell purity).
Additionally, we would expect the heralding efficiency to saturate to a constant value as the two-photon detuning is made large.
However, we see a clear peak in the heralding efficiency as the two-photon detuning is increased.
We believe this is related to the increased coupling power needed at large two-photon detunings.

In addition to a clear trend as the two-photon detuning is changed, for fixed two-photon detuning, we similarly see a peak in the heralding efficiency as the vapor cell temperature is altered.
This phenomenon has been previously observed \cite{davidson_bright_2021} and is attributed to the competing processes of an increase in directed collective emission, and decrease in idler photon transmission, as the vapor temperature, and therefore OD, is increased.

\begin{figure}[t]
    \centering
    \includegraphics[width=\linewidth]{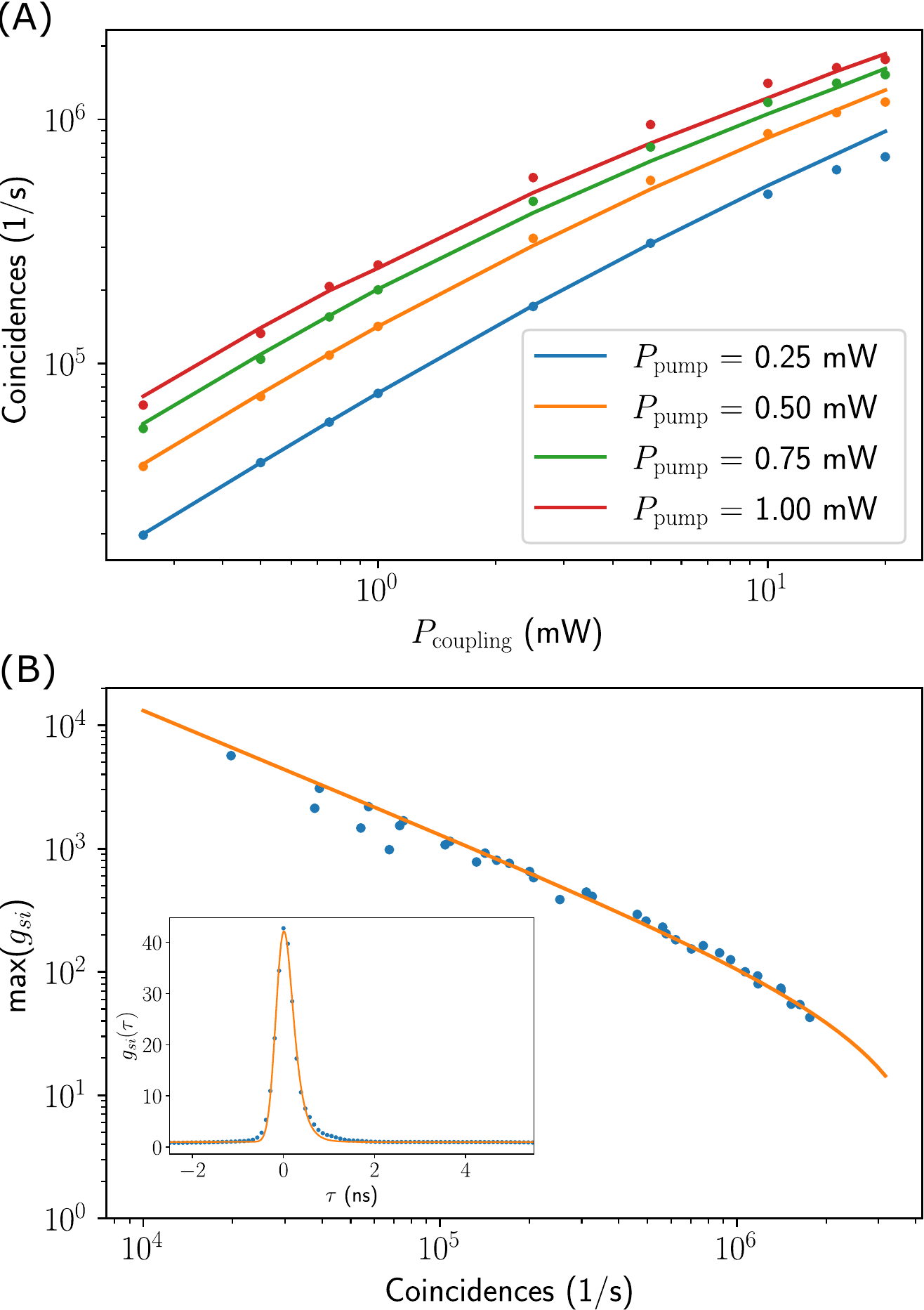}
    \caption{Scaling behavior of the source with pump and coupling power.
    (A) displays measured signal-idler coincidence rate as a function of coupling power, for various pump powers.
    We use a fit that accounts for the finite detector dead times.
    From a linear fit to the data we calculate a measured scaling constant of the coincidence rate $\approx 3\times10^5\, /s/\mathrm{mW}^{2}$ ($\approx 6\times10^5 \, /s/\mathrm{mW}^{2}$ when accounting for detection efficiencies).
    (B) shows the peak value of the signal-idler cross-correlation, $g_{si}$, as a function of the coincidence rate.
    Inset displays a typical $g_{si}$ curve, using $100$-ps bins.
    Orange line is a fit, taking into account the finite detector dead time in the system.
    In both plots error bars, arising from statistical uncertainties, are smaller than the data points.}
    \label{fig:power}
\end{figure}

For the remainder of this paper we operate the source with a two-photon detuning of $\delta\approx -2\pi\times 2400$ MHz and a cell temperature $\approx 93^\circ$ C, where the maximum measured heralding efficiency is, $\approx 16$ \% ($\approx 24$ \% corrected for idler detection efficiency).
Here, the resonant OD of the cell for the idler photons is $\approx 9$.

Next, we investigate scaling properties of the source with the coupling and pump powers.
Here, we again only measure the $\ket{VV}$ mode of the source.
In Fig. \ref{fig:power}(A) we show how the coincidence rate varies with the pump and coupling powers.
For low pump and coupling power we see a near-linear scaling in the coincidence rate as function of power.
At high power we observe a saturation in the coincidence rate for increasing power.
This is in part due to the finite dead time of the detectors ($\approx 20$ ns for both signal and idler), which we account for in the fit shown in the figure.
However, we have repeated the measurements shown in Fig. \ref{fig:power}(A), using neutral density filters on both the signal and idler paths to reduce this issue, and we still observe deviation from linearity at higher coupling powers.
We attribute this to saturation of the atomic medium.
From the fit in the linear regime, we find a measured scaling constant of $\approx 3\times10^5\, /s/\mathrm{mW}^{2}$ ($\approx 6\times10^5 \, /s/\mathrm{mW}^{2}$ when accounting for detection efficiencies).

Similar to other pair sources, as the pair production rate is increased, we expect the signal-idler cross-correlation function, $g_{si}$ to decrease.
We show the scaling of the maximum value of $g_{si}$ with coincidences in Fig. \ref{fig:power}(B), with a typical curve of $g_{si}$ as a function of the delay time between detections, $\tau$, shown inset.
In theory $g_{si}\propto 1/\mathrm{Coincidences}$.
We observe this inverse scaling for low coincidence rates.
However, for high coincidence rates we see a deviation from this behavior.
We attribute this to the finite detector dead time, as this is not seen when the measurement is repeated using neutral density filters on the signal and idler arms.
We take this detector saturation into account in the model in the figure.

At the maximum used powers of $1$ and $20$ mW, for the pump and coupling beams respectively (used for the remainder of the paper), the measured $\ket{VV}$ mode coincidence rate is $\approx 1.7\times10^6/s$ with $g_{si}\approx 40$, corresponding to a $\ket{VV}$ mode rate of $\approx 5\times10^6/s$, when correcting for the detector quantum efficiency and dead time saturation of the detectors.

Based on the $g_{si}$ curve, shown in the inset of Fig. \ref{fig:power}(B), we estimate the biphoton linewidth to be under $2\pi\times1$ GHz (after deconvolving the finite response time of the detectors), similar to that in the ladder-type FWM systems where all velocity classes participate in the collective excitation \cite{davidson_bright_2021}.
This value is corroborated by stimulated FWM biphoton measurements, similar to those in \cite{jeong_stimulated_2019}.
We attribute the biphoton bandwidth to the convolution of the Doppler broadened emission and the absorption of the idler as it propagates through the cell.
We note that this bandwidth is comparable to those demonstrated by various existing warm atom quantum memories \cite{finkelstein_fast_2018, buser_single-photon_2022}, opening the door for an all warm-atom based telecom compatible quantum repeater.

\begin{figure}
    \centering
    \includegraphics[width=\linewidth]{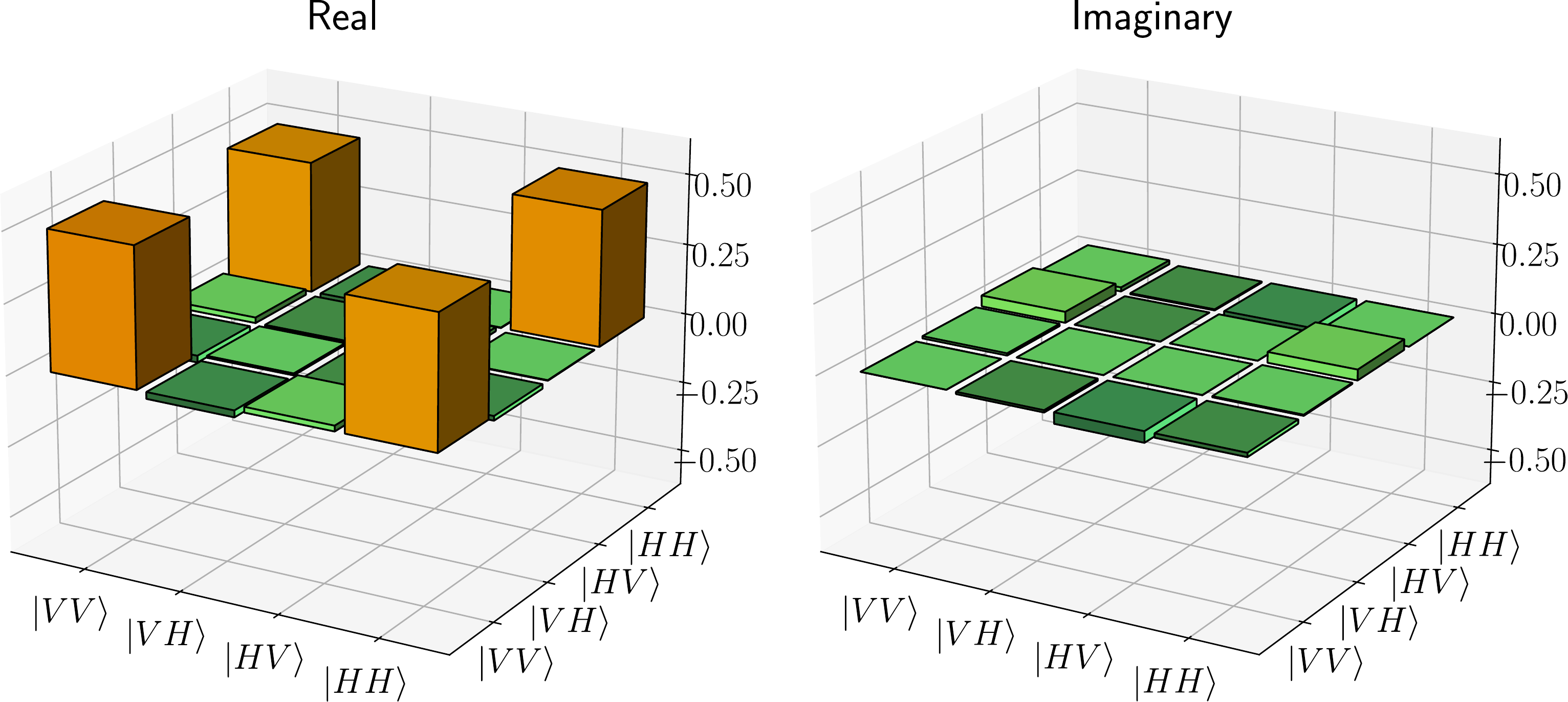}
    \caption{Real and imaginary parts of the maximum-likelihood density matrix reconstruction of the signal-idler state.
    Lower bound on the fidelity to the expected $\ket{\Phi_+}$ Bell state is $95$\%.}
    \label{fig:entanglement}
\end{figure}

Finally, with the pumping scheme used, and the rubidium Zeeman structure, we expect the source to produce pairs which are $\ket{\Phi_+}=\frac{1}{\sqrt{2}}\left(\ket{HH}+\ket{VV}\right)$ entangled \cite{willis_photon_2011}.
We find that the dichroic used to separate the signal and idler photons, as well as mirrors used to couple the signal and idler photons into their respective fibers, adds an arbitrary but stable phase shift between the $\ket{HH}$ and $\ket{VV}$ modes.
We tune the retardance of the LCR to compensate these phase shifts, and recover the $\ket{\Phi_+}$ state.
To verify the entangled state produced after this operation, we perform two-photon tomography.
We use neutral density filters with $OD\approx1$ on both the signal and idler paths to ensure there are no issues with detector saturation during the tomography.
Using the maximum likelihood method discussed in \cite{james_measurement_2001} we reconstruct the density matrix for the two-photon state, shown in Fig. \ref{fig:entanglement}.
From the reconstructed density matrix we place a lower bound fidelity to the $\ket{\Phi_+}$ Bell state of $95$\%\footnote{Lower bound is the minimum value attained when varying the minimization solver and initial conditions}, for an entangled pair rate greater than $10^7\, /s$.

\section{Conclusion}

The development of large-scale quantum networks relies on the development of practical entanglement sources that can satisfy multiple conditions such as a high pair generation rate, narrow linewidth, high fidelity and wavelength compatibility with telecom and atomic systems.
We have theoretically and experimentally investigated a new operating regime for warm-atom FWM entangled photon sources in the diamond configuration that can satisfy all the above conditions simultaneously.
Under these conditions, we are able to simultaneously address all the velocity groups within a vapor, enabling us to achieve an in-fiber entangled pair absolute brightness greater than $10^7/s$.
Given the sub GHz bandwidth of the biphotons produced the corresponding spectral brightness is $\approx10^4/s/\mathrm{MHz}$.
To the best of our knowledge, this is the highest absolute and spectral brightness achieved for a warm-atom entangled photon source.
We have demonstrated that the bichromatic photon pairs ($1324$ and $795$ nm) are well correlated, $g_{si} \geq 40$, and from maximum likelihood estimation we place a lower bound on the fidelity to the $\ket{\Phi_+}$ Bell state of $95$\%.
Given that our source produces entangled pairs that are compatible both with telecom infrastructure and existing quantum memories, it has applications for quantum repeating, distributed quantum processing, and quantum enhanced sensor networks. 
Additionally, the relative simplicity of the source should allow for the development of robust devices for integrating into existing telecom infrastructure.

\section*{Acknowledgments}
We thank the Qunnect Inc. team, especially Noel Goddard, for their help with manuscript preparation. 
This work is supported by the U.S. Department of Energy under {award no. DE-SC0021556.

\bibliography{Source_Paper}

\begin{thebibliography}{27}%
\makeatletter
\providecommand \@ifxundefined [1]{%
 \@ifx{#1\undefined}
}%
\providecommand \@ifnum [1]{%
 \ifnum #1\expandafter \@firstoftwo
 \else \expandafter \@secondoftwo
 \fi
}%
\providecommand \@ifx [1]{%
 \ifx #1\expandafter \@firstoftwo
 \else \expandafter \@secondoftwo
 \fi
}%
\providecommand \natexlab [1]{#1}%
\providecommand \enquote  [1]{``#1''}%
\providecommand \bibnamefont  [1]{#1}%
\providecommand \bibfnamefont [1]{#1}%
\providecommand \citenamefont [1]{#1}%
\providecommand \href@noop [0]{\@secondoftwo}%
\providecommand \href [0]{\begingroup \@sanitize@url \@href}%
\providecommand \@href[1]{\@@startlink{#1}\@@href}%
\providecommand \@@href[1]{\endgroup#1\@@endlink}%
\providecommand \@sanitize@url [0]{\catcode `\\12\catcode `\$12\catcode
  `\&12\catcode `\#12\catcode `\^12\catcode `\_12\catcode `\%12\relax}%
\providecommand \@@startlink[1]{}%
\providecommand \@@endlink[0]{}%
\providecommand \url  [0]{\begingroup\@sanitize@url \@url }%
\providecommand \@url [1]{\endgroup\@href {#1}{\urlprefix }}%
\providecommand \urlprefix  [0]{URL }%
\providecommand \Eprint [0]{\href }%
\providecommand \doibase [0]{https://doi.org/}%
\providecommand \selectlanguage [0]{\@gobble}%
\providecommand \bibinfo  [0]{\@secondoftwo}%
\providecommand \bibfield  [0]{\@secondoftwo}%
\providecommand \translation [1]{[#1]}%
\providecommand \BibitemOpen [0]{}%
\providecommand \bibitemStop [0]{}%
\providecommand \bibitemNoStop [0]{.\EOS\space}%
\providecommand \EOS [0]{\spacefactor3000\relax}%
\providecommand \BibitemShut  [1]{\csname bibitem#1\endcsname}%
\let\auto@bib@innerbib\@empty
\bibitem [{\citenamefont {Monroe}\ \emph {et~al.}(2014)\citenamefont {Monroe},
  \citenamefont {Raussendorf}, \citenamefont {Ruthven}, \citenamefont {Brown},
  \citenamefont {Maunz}, \citenamefont {Duan},\ and\ \citenamefont
  {Kim}}]{monroe_large-scale_2014}%
  \BibitemOpen
  \bibfield  {author} {\bibinfo {author} {\bibfnamefont {C.}~\bibnamefont
  {Monroe}}, \bibinfo {author} {\bibfnamefont {R.}~\bibnamefont {Raussendorf}},
  \bibinfo {author} {\bibfnamefont {A.}~\bibnamefont {Ruthven}}, \bibinfo
  {author} {\bibfnamefont {K.~R.}\ \bibnamefont {Brown}}, \bibinfo {author}
  {\bibfnamefont {P.}~\bibnamefont {Maunz}}, \bibinfo {author} {\bibfnamefont
  {L.-M.}\ \bibnamefont {Duan}},\ and\ \bibinfo {author} {\bibfnamefont
  {J.}~\bibnamefont {Kim}},\ }\bibfield  {title} {\bibinfo {title} {Large-scale
  modular quantum-computer architecture with atomic memory and photonic
  interconnects},\ }\href@noop {} {\bibfield  {journal} {\bibinfo  {journal}
  {Phys. Rev. A}\ }\textbf {\bibinfo {volume} {89}} (\bibinfo {year}
  {2014})}\BibitemShut {NoStop}%
\bibitem [{\citenamefont {Eldredge}\ \emph {et~al.}(2018)\citenamefont
  {Eldredge}, \citenamefont {Foss-Feig}, \citenamefont {Gross}, \citenamefont
  {Rolston},\ and\ \citenamefont {Gorshkov}}]{eldredge_optimal_2018}%
  \BibitemOpen
  \bibfield  {author} {\bibinfo {author} {\bibfnamefont {Z.}~\bibnamefont
  {Eldredge}}, \bibinfo {author} {\bibfnamefont {M.}~\bibnamefont {Foss-Feig}},
  \bibinfo {author} {\bibfnamefont {J.~A.}\ \bibnamefont {Gross}}, \bibinfo
  {author} {\bibfnamefont {S.~L.}\ \bibnamefont {Rolston}},\ and\ \bibinfo
  {author} {\bibfnamefont {A.~V.}\ \bibnamefont {Gorshkov}},\ }\bibfield
  {title} {\bibinfo {title} {Optimal and secure measurement protocols for
  quantum sensor networks},\ }\href@noop {} {\bibfield  {journal} {\bibinfo
  {journal} {Phys. Rev. A}\ }\textbf {\bibinfo {volume} {97}} (\bibinfo {year}
  {2018})}\BibitemShut {NoStop}%
\bibitem [{\citenamefont {Ekert}(1991)}]{ekert_quantum_1991}%
  \BibitemOpen
  \bibfield  {author} {\bibinfo {author} {\bibfnamefont {A.~K.}\ \bibnamefont
  {Ekert}},\ }\bibfield  {title} {\bibinfo {title} {Quantum cryptography based
  on bell's theorem},\ }\href@noop {} {\bibfield  {journal} {\bibinfo
  {journal} {Phys. Rev. Lett.}\ }\textbf {\bibinfo {volume} {67}},\ \bibinfo
  {pages} {661} (\bibinfo {year} {1991})}\BibitemShut {NoStop}%
\bibitem [{\citenamefont {Duan}\ \emph {et~al.}(2001)\citenamefont {Duan},
  \citenamefont {Lukin}, \citenamefont {Cirac},\ and\ \citenamefont
  {Zoller}}]{Duan2001}%
  \BibitemOpen
  \bibfield  {author} {\bibinfo {author} {\bibfnamefont {L.~M.}\ \bibnamefont
  {Duan}}, \bibinfo {author} {\bibfnamefont {M.~D.}\ \bibnamefont {Lukin}},
  \bibinfo {author} {\bibfnamefont {J.~I.}\ \bibnamefont {Cirac}},\ and\
  \bibinfo {author} {\bibfnamefont {P.}~\bibnamefont {Zoller}},\ }\bibfield
  {title} {\bibinfo {title} {Long-distance quantum communication with atomic
  ensembles and linear optics},\ }\href@noop {} {\bibfield  {journal} {\bibinfo
   {journal} {Nature}\ }\textbf {\bibinfo {volume} {414}},\ \bibinfo {pages}
  {413} (\bibinfo {year} {2001})}\BibitemShut {NoStop}%
\bibitem [{\citenamefont {Sangouard}\ \emph {et~al.}(2011)\citenamefont
  {Sangouard}, \citenamefont {Simon}, \citenamefont {de~Riedmatten},\ and\
  \citenamefont {Gisin}}]{sangouard_quantum_2009}%
  \BibitemOpen
  \bibfield  {author} {\bibinfo {author} {\bibfnamefont {N.}~\bibnamefont
  {Sangouard}}, \bibinfo {author} {\bibfnamefont {C.}~\bibnamefont {Simon}},
  \bibinfo {author} {\bibfnamefont {H.}~\bibnamefont {de~Riedmatten}},\ and\
  \bibinfo {author} {\bibfnamefont {N.}~\bibnamefont {Gisin}},\ }\bibfield
  {title} {\bibinfo {title} {Quantum repeaters based on atomic ensembles and
  linear optics},\ }\href@noop {} {\bibfield  {journal} {\bibinfo  {journal}
  {Rev. Mod. Phys.}\ }\textbf {\bibinfo {volume} {83}},\ \bibinfo {pages} {33}
  (\bibinfo {year} {2011})}\BibitemShut {NoStop}%
\bibitem [{\citenamefont {Davidson}\ \emph {et~al.}(2021)\citenamefont
  {Davidson}, \citenamefont {Finkelstein}, \citenamefont {Poem},\ and\
  \citenamefont {Firstenberg}}]{davidson_bright_2021}%
  \BibitemOpen
  \bibfield  {author} {\bibinfo {author} {\bibfnamefont {O.}~\bibnamefont
  {Davidson}}, \bibinfo {author} {\bibfnamefont {R.}~\bibnamefont
  {Finkelstein}}, \bibinfo {author} {\bibfnamefont {E.}~\bibnamefont {Poem}},\
  and\ \bibinfo {author} {\bibfnamefont {O.}~\bibnamefont {Firstenberg}},\
  }\bibfield  {title} {\bibinfo {title} {Bright multiplexed source of
  indistinguishable single photons with tunable {GHz-bandwidth} at room
  temperature},\ }\href@noop {} {\bibfield  {journal} {\bibinfo  {journal} {New
  J. Phys.}\ }\textbf {\bibinfo {volume} {23}},\ \bibinfo {pages} {073050}
  (\bibinfo {year} {2021})}\BibitemShut {NoStop}%
\bibitem [{\citenamefont {Lee}\ \emph {et~al.}(2016)\citenamefont {Lee},
  \citenamefont {Lee}, \citenamefont {Kim},\ and\ \citenamefont
  {Moon}}]{lee_highly_2016}%
  \BibitemOpen
  \bibfield  {author} {\bibinfo {author} {\bibfnamefont {Y.-S.}\ \bibnamefont
  {Lee}}, \bibinfo {author} {\bibfnamefont {S.~M.}\ \bibnamefont {Lee}},
  \bibinfo {author} {\bibfnamefont {H.}~\bibnamefont {Kim}},\ and\ \bibinfo
  {author} {\bibfnamefont {H.~S.}\ \bibnamefont {Moon}},\ }\bibfield  {title}
  {\bibinfo {title} {Highly bright photon-pair generation in doppler-broadened
  ladder-type atomic system},\ }\href@noop {} {\bibfield  {journal} {\bibinfo
  {journal} {Opt. Express}\ }\textbf {\bibinfo {volume} {24}},\ \bibinfo
  {pages} {28083} (\bibinfo {year} {2016})}\BibitemShut {NoStop}%
\bibitem [{\citenamefont {Davidson}\ \emph {et~al.}()\citenamefont {Davidson},
  \citenamefont {Yogev}, \citenamefont {Poem},\ and\ \citenamefont
  {Firstenberg}}]{davidson_bright_2023}%
  \BibitemOpen
  \bibfield  {author} {\bibinfo {author} {\bibfnamefont {O.}~\bibnamefont
  {Davidson}}, \bibinfo {author} {\bibfnamefont {O.}~\bibnamefont {Yogev}},
  \bibinfo {author} {\bibfnamefont {E.}~\bibnamefont {Poem}},\ and\ \bibinfo
  {author} {\bibfnamefont {O.}~\bibnamefont {Firstenberg}},\ }\href@noop {}
  {\bibinfo {title} {Bright, low-noise source of single photons at 780 nm with
  improved phase-matching in rubidium vapor}},\ \Eprint
  {https://arxiv.org/abs/arXiv:2301.06049} {arXiv:2301.06049} \BibitemShut
  {NoStop}%
\bibitem [{\citenamefont {Park}\ \emph {et~al.}(2019)\citenamefont {Park},
  \citenamefont {Kim},\ and\ \citenamefont
  {Moon}}]{park_polarization-entangled_2019}%
  \BibitemOpen
  \bibfield  {author} {\bibinfo {author} {\bibfnamefont {J.}~\bibnamefont
  {Park}}, \bibinfo {author} {\bibfnamefont {H.}~\bibnamefont {Kim}},\ and\
  \bibinfo {author} {\bibfnamefont {H.~S.}\ \bibnamefont {Moon}},\ }\bibfield
  {title} {\bibinfo {title} {Polarization-entangled photons from a warm atomic
  ensemble using a sagnac interferometer},\ }\href@noop {} {\bibfield
  {journal} {\bibinfo  {journal} {Phys. Rev. Lett.}\ }\textbf {\bibinfo
  {volume} {122}},\ \bibinfo {pages} {143601} (\bibinfo {year}
  {2019})}\BibitemShut {NoStop}%
\bibitem [{\citenamefont {Park}\ \emph {et~al.}(2021)\citenamefont {Park},
  \citenamefont {Bae}, \citenamefont {Kim},\ and\ \citenamefont
  {Moon}}]{park_direct_2021}%
  \BibitemOpen
  \bibfield  {author} {\bibinfo {author} {\bibfnamefont {J.}~\bibnamefont
  {Park}}, \bibinfo {author} {\bibfnamefont {J.}~\bibnamefont {Bae}}, \bibinfo
  {author} {\bibfnamefont {H.}~\bibnamefont {Kim}},\ and\ \bibinfo {author}
  {\bibfnamefont {H.~S.}\ \bibnamefont {Moon}},\ }\bibfield  {title} {\bibinfo
  {title} {Direct generation of polarization-entangled photons from warm atomic
  ensemble},\ }\href@noop {} {\bibfield  {journal} {\bibinfo  {journal} {Appl.
  Phys. Lett.}\ }\textbf {\bibinfo {volume} {119}},\ \bibinfo {pages} {074001}
  (\bibinfo {year} {2021})}\BibitemShut {NoStop}%
\bibitem [{\citenamefont {Ding}\ \emph {et~al.}(2012)\citenamefont {Ding},
  \citenamefont {Zhou}, \citenamefont {Shi}, \citenamefont {Zou},\ and\
  \citenamefont {Guo}}]{ding_generation_2012}%
  \BibitemOpen
  \bibfield  {author} {\bibinfo {author} {\bibfnamefont {D.-S.}\ \bibnamefont
  {Ding}}, \bibinfo {author} {\bibfnamefont {Z.-Y.}\ \bibnamefont {Zhou}},
  \bibinfo {author} {\bibfnamefont {B.-S.}\ \bibnamefont {Shi}}, \bibinfo
  {author} {\bibfnamefont {X.-B.}\ \bibnamefont {Zou}},\ and\ \bibinfo {author}
  {\bibfnamefont {G.-C.}\ \bibnamefont {Guo}},\ }\bibfield  {title} {\bibinfo
  {title} {Generation of non-classical correlated photon pairs via a
  ladder-type atomic configuration: theory and experiment},\ }\href@noop {}
  {\bibfield  {journal} {\bibinfo  {journal} {Opt. Express}\ }\textbf {\bibinfo
  {volume} {20}},\ \bibinfo {pages} {11433} (\bibinfo {year}
  {2012})}\BibitemShut {NoStop}%
\bibitem [{\citenamefont {Chen}\ \emph {et~al.}(2022)\citenamefont {Chen},
  \citenamefont {Hsu}, \citenamefont {Huang}, \citenamefont {Hsiao},
  \citenamefont {Huang}, \citenamefont {Chen}, \citenamefont {Chuu},
  \citenamefont {Chen}, \citenamefont {Chen},\ and\ \citenamefont
  {Yu}}]{chen_room-temperature_2022}%
  \BibitemOpen
  \bibfield  {author} {\bibinfo {author} {\bibfnamefont {J.-M.}\ \bibnamefont
  {Chen}}, \bibinfo {author} {\bibfnamefont {C.-Y.}\ \bibnamefont {Hsu}},
  \bibinfo {author} {\bibfnamefont {W.-K.}\ \bibnamefont {Huang}}, \bibinfo
  {author} {\bibfnamefont {S.-S.}\ \bibnamefont {Hsiao}}, \bibinfo {author}
  {\bibfnamefont {F.-C.}\ \bibnamefont {Huang}}, \bibinfo {author}
  {\bibfnamefont {Y.-H.}\ \bibnamefont {Chen}}, \bibinfo {author}
  {\bibfnamefont {C.-S.}\ \bibnamefont {Chuu}}, \bibinfo {author}
  {\bibfnamefont {Y.-C.}\ \bibnamefont {Chen}}, \bibinfo {author}
  {\bibfnamefont {Y.-F.}\ \bibnamefont {Chen}},\ and\ \bibinfo {author}
  {\bibfnamefont {I.~A.}\ \bibnamefont {Yu}},\ }\bibfield  {title} {\bibinfo
  {title} {Room-temperature biphoton source with a spectral brightness near the
  ultimate limit},\ }\href@noop {} {\bibfield  {journal} {\bibinfo  {journal}
  {Phys. Rev. Res.}\ }\textbf {\bibinfo {volume} {4}} (\bibinfo {year}
  {2022})}\BibitemShut {NoStop}%
\bibitem [{\citenamefont {Hsu}\ \emph {et~al.}(2021)\citenamefont {Hsu},
  \citenamefont {Wang}, \citenamefont {Chen}, \citenamefont {Huang},
  \citenamefont {Ke}, \citenamefont {Huang}, \citenamefont {Hung},
  \citenamefont {Chao}, \citenamefont {Hsiao}, \citenamefont {Chen},
  \citenamefont {Chuu}, \citenamefont {Chen}, \citenamefont {Chen},\ and\
  \citenamefont {Yu}}]{hsu_generation_2021}%
  \BibitemOpen
  \bibfield  {author} {\bibinfo {author} {\bibfnamefont {C.-Y.}\ \bibnamefont
  {Hsu}}, \bibinfo {author} {\bibfnamefont {Y.-S.}\ \bibnamefont {Wang}},
  \bibinfo {author} {\bibfnamefont {J.-M.}\ \bibnamefont {Chen}}, \bibinfo
  {author} {\bibfnamefont {F.-C.}\ \bibnamefont {Huang}}, \bibinfo {author}
  {\bibfnamefont {Y.-T.}\ \bibnamefont {Ke}}, \bibinfo {author} {\bibfnamefont
  {E.~K.}\ \bibnamefont {Huang}}, \bibinfo {author} {\bibfnamefont
  {W.}~\bibnamefont {Hung}}, \bibinfo {author} {\bibfnamefont {K.-L.}\
  \bibnamefont {Chao}}, \bibinfo {author} {\bibfnamefont {S.-S.}\ \bibnamefont
  {Hsiao}}, \bibinfo {author} {\bibfnamefont {Y.-H.}\ \bibnamefont {Chen}},
  \bibinfo {author} {\bibfnamefont {C.-S.}\ \bibnamefont {Chuu}}, \bibinfo
  {author} {\bibfnamefont {Y.-C.}\ \bibnamefont {Chen}}, \bibinfo {author}
  {\bibfnamefont {Y.-F.}\ \bibnamefont {Chen}},\ and\ \bibinfo {author}
  {\bibfnamefont {I.~A.}\ \bibnamefont {Yu}},\ }\bibfield  {title} {\bibinfo
  {title} {Generation of {sub-MHz} and spectrally-bright biphotons from hot
  atomic vapors with a phase mismatch-free scheme},\ }\href@noop {} {\bibfield
  {journal} {\bibinfo  {journal} {Opt. Express}\ }\textbf {\bibinfo {volume}
  {29}},\ \bibinfo {pages} {4632} (\bibinfo {year} {2021})}\BibitemShut
  {NoStop}%
\bibitem [{\citenamefont {Shu}\ \emph {et~al.}(2016)\citenamefont {Shu},
  \citenamefont {Chen}, \citenamefont {Chow}, \citenamefont {Zhu},
  \citenamefont {Xiao}, \citenamefont {Loy},\ and\ \citenamefont
  {Du}}]{shu_subnatural-linewidth_2016}%
  \BibitemOpen
  \bibfield  {author} {\bibinfo {author} {\bibfnamefont {C.}~\bibnamefont
  {Shu}}, \bibinfo {author} {\bibfnamefont {P.}~\bibnamefont {Chen}}, \bibinfo
  {author} {\bibfnamefont {T.~K.~A.}\ \bibnamefont {Chow}}, \bibinfo {author}
  {\bibfnamefont {L.}~\bibnamefont {Zhu}}, \bibinfo {author} {\bibfnamefont
  {Y.}~\bibnamefont {Xiao}}, \bibinfo {author} {\bibfnamefont {M.~M.~T.}\
  \bibnamefont {Loy}},\ and\ \bibinfo {author} {\bibfnamefont {S.}~\bibnamefont
  {Du}},\ }\bibfield  {title} {\bibinfo {title} {Subnatural-linewidth biphotons
  from a doppler-broadened hot atomic vapour cell},\ }\href@noop {} {\bibfield
  {journal} {\bibinfo  {journal} {Nat. Commun.}\ }\textbf {\bibinfo {volume}
  {7}},\ \bibinfo {pages} {12783} (\bibinfo {year} {2016})}\BibitemShut
  {NoStop}%
\bibitem [{\citenamefont {Zhu}\ \emph {et~al.}(2017)\citenamefont {Zhu},
  \citenamefont {Guo}, \citenamefont {Shu}, \citenamefont {Jeong},\ and\
  \citenamefont {Du}}]{zhu_bright_2017}%
  \BibitemOpen
  \bibfield  {author} {\bibinfo {author} {\bibfnamefont {L.}~\bibnamefont
  {Zhu}}, \bibinfo {author} {\bibfnamefont {X.}~\bibnamefont {Guo}}, \bibinfo
  {author} {\bibfnamefont {C.}~\bibnamefont {Shu}}, \bibinfo {author}
  {\bibfnamefont {H.}~\bibnamefont {Jeong}},\ and\ \bibinfo {author}
  {\bibfnamefont {S.}~\bibnamefont {Du}},\ }\bibfield  {title} {\bibinfo
  {title} {Bright narrowband biphoton generation from a hot rubidium atomic
  vapor cell},\ }\href@noop {} {\bibfield  {journal} {\bibinfo  {journal}
  {Appl. Phys. Lett.}\ }\textbf {\bibinfo {volume} {110}},\ \bibinfo {pages}
  {161101} (\bibinfo {year} {2017})}\BibitemShut {NoStop}%
\bibitem [{\citenamefont {Mika}\ and\ \citenamefont {Slodi{\v
  c}ka}(2020)}]{mika_high_2020}%
  \BibitemOpen
  \bibfield  {author} {\bibinfo {author} {\bibfnamefont {J.}~\bibnamefont
  {Mika}}\ and\ \bibinfo {author} {\bibfnamefont {L.}~\bibnamefont {Slodi{\v
  c}ka}},\ }\bibfield  {title} {\bibinfo {title} {High nonclassical
  correlations of large-bandwidth photon pairs generated in warm atomic
  vapor},\ }\href@noop {} {\bibfield  {journal} {\bibinfo  {journal} {J. Phys.
  B At. Mol. Opt. Phys.}\ }\textbf {\bibinfo {volume} {53}},\ \bibinfo {pages}
  {145501} (\bibinfo {year} {2020})}\BibitemShut {NoStop}%
\bibitem [{\citenamefont {Jeong}\ and\ \citenamefont
  {Moon}(2020)}]{jeong_temporal-_2020}%
  \BibitemOpen
  \bibfield  {author} {\bibinfo {author} {\bibfnamefont {T.}~\bibnamefont
  {Jeong}}\ and\ \bibinfo {author} {\bibfnamefont {H.~S.}\ \bibnamefont
  {Moon}},\ }\bibfield  {title} {\bibinfo {title} {Temporal- and
  spectral-property measurements of narrowband photon pairs from warm
  {double-$\Lambda$-type} atomic ensemble},\ }\href@noop {} {\bibfield
  {journal} {\bibinfo  {journal} {Opt. Express}\ }\textbf {\bibinfo {volume}
  {28}},\ \bibinfo {pages} {3985} (\bibinfo {year} {2020})}\BibitemShut
  {NoStop}%
\bibitem [{\citenamefont {Willis}\ \emph {et~al.}(2010)\citenamefont {Willis},
  \citenamefont {Becerra}, \citenamefont {Orozco},\ and\ \citenamefont
  {Rolston}}]{willis_correlated_2010}%
  \BibitemOpen
  \bibfield  {author} {\bibinfo {author} {\bibfnamefont {R.~T.}\ \bibnamefont
  {Willis}}, \bibinfo {author} {\bibfnamefont {F.~E.}\ \bibnamefont {Becerra}},
  \bibinfo {author} {\bibfnamefont {L.~A.}\ \bibnamefont {Orozco}},\ and\
  \bibinfo {author} {\bibfnamefont {S.~L.}\ \bibnamefont {Rolston}},\
  }\bibfield  {title} {\bibinfo {title} {Correlated photon pairs generated from
  a warm atomic ensemble},\ }\href@noop {} {\bibfield  {journal} {\bibinfo
  {journal} {Phys. Rev. A}\ }\textbf {\bibinfo {volume} {82}} (\bibinfo {year}
  {2010})}\BibitemShut {NoStop}%
\bibitem [{\citenamefont {Willis}\ \emph {et~al.}(2011)\citenamefont {Willis},
  \citenamefont {Becerra}, \citenamefont {Orozco},\ and\ \citenamefont
  {Rolston}}]{willis_photon_2011}%
  \BibitemOpen
  \bibfield  {author} {\bibinfo {author} {\bibfnamefont {R.~T.}\ \bibnamefont
  {Willis}}, \bibinfo {author} {\bibfnamefont {F.~E.}\ \bibnamefont {Becerra}},
  \bibinfo {author} {\bibfnamefont {L.~A.}\ \bibnamefont {Orozco}},\ and\
  \bibinfo {author} {\bibfnamefont {S.~L.}\ \bibnamefont {Rolston}},\
  }\bibfield  {title} {\bibinfo {title} {Photon statistics and polarization
  correlations at telecommunications wavelengths from a warm atomic ensemble},\
  }\href@noop {} {\bibfield  {journal} {\bibinfo  {journal} {Opt. Express}\
  }\textbf {\bibinfo {volume} {19}},\ \bibinfo {pages} {14632} (\bibinfo {year}
  {2011})}\BibitemShut {NoStop}%
\bibitem [{\citenamefont {Wang}\ \emph {et~al.}(2022)\citenamefont {Wang},
  \citenamefont {Craddock}, \citenamefont {Sekelsky}, \citenamefont {Flament},\
  and\ \citenamefont {Namazi}}]{wang_field-deployable_2022}%
  \BibitemOpen
  \bibfield  {author} {\bibinfo {author} {\bibfnamefont {Y.}~\bibnamefont
  {Wang}}, \bibinfo {author} {\bibfnamefont {A.~N.}\ \bibnamefont {Craddock}},
  \bibinfo {author} {\bibfnamefont {R.}~\bibnamefont {Sekelsky}}, \bibinfo
  {author} {\bibfnamefont {M.}~\bibnamefont {Flament}},\ and\ \bibinfo {author}
  {\bibfnamefont {M.}~\bibnamefont {Namazi}},\ }\bibfield  {title} {\bibinfo
  {title} {Field-deployable quantum memory for quantum networking},\
  }\href@noop {} {\bibfield  {journal} {\bibinfo  {journal} {Phys. Rev. Appl.}\
  }\textbf {\bibinfo {volume} {18}} (\bibinfo {year} {2022})}\BibitemShut
  {NoStop}%
\bibitem [{\citenamefont {Saffman}\ and\ \citenamefont
  {Walker}(2002)}]{saffman_creating_2002}%
  \BibitemOpen
  \bibfield  {author} {\bibinfo {author} {\bibfnamefont {M.}~\bibnamefont
  {Saffman}}\ and\ \bibinfo {author} {\bibfnamefont {T.~G.}\ \bibnamefont
  {Walker}},\ }\bibfield  {title} {\bibinfo {title} {Creating single-atom and
  single-photon sources from entangled atomic ensembles},\ }\href@noop {}
  {\bibfield  {journal} {\bibinfo  {journal} {Phys. Rev. A}\ }\textbf {\bibinfo
  {volume} {66}} (\bibinfo {year} {2002})}\BibitemShut {NoStop}%
\bibitem [{\citenamefont {P{\'e}rez~Galv{\'a}n}\ \emph
  {et~al.}(2008)\citenamefont {P{\'e}rez~Galv{\'a}n}, \citenamefont {Zhao},\
  and\ \citenamefont {Orozco}}]{perez_galvan_measurement_2008}%
  \BibitemOpen
  \bibfield  {author} {\bibinfo {author} {\bibfnamefont {A.}~\bibnamefont
  {P{\'e}rez~Galv{\'a}n}}, \bibinfo {author} {\bibfnamefont {Y.}~\bibnamefont
  {Zhao}},\ and\ \bibinfo {author} {\bibfnamefont {L.~A.}\ \bibnamefont
  {Orozco}},\ }\bibfield  {title} {\bibinfo {title} {Measurement of the
  hyperfine splitting of the $6{S}_{1/2}$ level in rubidium},\ }\href@noop {}
  {\bibfield  {journal} {\bibinfo  {journal} {Phys. Rev. A}\ }\textbf {\bibinfo
  {volume} {78}} (\bibinfo {year} {2008})}\BibitemShut {NoStop}%
\bibitem [{\citenamefont {Jeong}\ \emph {et~al.}(2019)\citenamefont {Jeong},
  \citenamefont {Park},\ and\ \citenamefont {Moon}}]{jeong_stimulated_2019}%
  \BibitemOpen
  \bibfield  {author} {\bibinfo {author} {\bibfnamefont {T.}~\bibnamefont
  {Jeong}}, \bibinfo {author} {\bibfnamefont {J.}~\bibnamefont {Park}},\ and\
  \bibinfo {author} {\bibfnamefont {H.~S.}\ \bibnamefont {Moon}},\ }\bibfield
  {title} {\bibinfo {title} {Stimulated measurement of spontaneous four-wave
  mixing from a warm atomic ensemble},\ }\href@noop {} {\bibfield  {journal}
  {\bibinfo  {journal} {Phys. Rev. A (Coll. Park.)}\ }\textbf {\bibinfo
  {volume} {100}} (\bibinfo {year} {2019})}\BibitemShut {NoStop}%
\bibitem [{\citenamefont {Finkelstein}\ \emph {et~al.}(2018)\citenamefont
  {Finkelstein}, \citenamefont {Poem}, \citenamefont {Michel}, \citenamefont
  {Lahad},\ and\ \citenamefont {Firstenberg}}]{finkelstein_fast_2018}%
  \BibitemOpen
  \bibfield  {author} {\bibinfo {author} {\bibfnamefont {R.}~\bibnamefont
  {Finkelstein}}, \bibinfo {author} {\bibfnamefont {E.}~\bibnamefont {Poem}},
  \bibinfo {author} {\bibfnamefont {O.}~\bibnamefont {Michel}}, \bibinfo
  {author} {\bibfnamefont {O.}~\bibnamefont {Lahad}},\ and\ \bibinfo {author}
  {\bibfnamefont {O.}~\bibnamefont {Firstenberg}},\ }\bibfield  {title}
  {\bibinfo {title} {Fast, noise-free memory for photon synchronization at room
  temperature},\ }\href@noop {} {\bibfield  {journal} {\bibinfo  {journal}
  {Sci. Adv.}\ }\textbf {\bibinfo {volume} {4}},\ \bibinfo {pages} {eaap8598}
  (\bibinfo {year} {2018})}\BibitemShut {NoStop}%
\bibitem [{\citenamefont {Buser}\ \emph {et~al.}(2022)\citenamefont {Buser},
  \citenamefont {Mottola}, \citenamefont {Cotting}, \citenamefont {Wolters},\
  and\ \citenamefont {Treutlein}}]{buser_single-photon_2022}%
  \BibitemOpen
  \bibfield  {author} {\bibinfo {author} {\bibfnamefont {G.}~\bibnamefont
  {Buser}}, \bibinfo {author} {\bibfnamefont {R.}~\bibnamefont {Mottola}},
  \bibinfo {author} {\bibfnamefont {B.}~\bibnamefont {Cotting}}, \bibinfo
  {author} {\bibfnamefont {J.}~\bibnamefont {Wolters}},\ and\ \bibinfo {author}
  {\bibfnamefont {P.}~\bibnamefont {Treutlein}},\ }\bibfield  {title} {\bibinfo
  {title} {Single-photon storage in a ground-state vapor cell quantum memory},\
  }\href@noop {} {\bibfield  {journal} {\bibinfo  {journal} {PRX quantum}\
  }\textbf {\bibinfo {volume} {3}} (\bibinfo {year} {2022})}\BibitemShut
  {NoStop}%
\bibitem [{\citenamefont {James}\ \emph {et~al.}(2001)\citenamefont {James},
  \citenamefont {Kwiat}, \citenamefont {Munro},\ and\ \citenamefont
  {White}}]{james_measurement_2001}%
  \BibitemOpen
  \bibfield  {author} {\bibinfo {author} {\bibfnamefont {D.~F.~V.}\
  \bibnamefont {James}}, \bibinfo {author} {\bibfnamefont {P.~G.}\ \bibnamefont
  {Kwiat}}, \bibinfo {author} {\bibfnamefont {W.~J.}\ \bibnamefont {Munro}},\
  and\ \bibinfo {author} {\bibfnamefont {A.~G.}\ \bibnamefont {White}},\
  }\bibfield  {title} {\bibinfo {title} {Measurement of qubits},\ }\href@noop
  {} {\bibfield  {journal} {\bibinfo  {journal} {Phys. Rev. A}\ }\textbf
  {\bibinfo {volume} {64}} (\bibinfo {year} {2001})}\BibitemShut {NoStop}%
\bibitem [{Note1()}]{Note1}%
  \BibitemOpen
  \bibinfo {note} {Lower bound is the minimum value attained when varying the
  minimization solver and initial conditions}\BibitemShut {NoStop}%
\end{thebibliography}%

\end{document}